\documentclass[a4paper,12pt]{article}
\usepackage[cp1251]{inputenc}
\usepackage{graphics}
\usepackage{amssymb,amsmath}
\usepackage[english]{babel}
\usepackage{indentfirst}
\begin{document}

\bigskip
\centerline {\bf SOLAR AND SOLAR-TYPE STARS ATMOSPHERE'S ACTIVITY}
\centerline {\bf AT 11-YEAR AND QUASI-BIENNIAL TIME SCALES}

\bigskip

\centerline {E.A.Bruevich $^{a}$, E.V. Kononovich $^{b}$}

\bigskip

\centerline {Sternberg Astronomical Institute, Moscow, Russia}\

\centerline
{E-mail:$^a${red-field@yandex.ru},$^b${konon@sai.msu.ru}}\

\bigskip

{\bf Abstract.} We studied new simultaneous observations of the flux
variations of the photospheric and chromospheric emissions of 33
solar-type stars and Sun during the HK-project that were conducted
over the past 20 years. In addition to the known cyclic
chromospheric emission variations of stars at the 11-year time
scale, which were discovered at the Mount Wilson Observatory, we
found a recurrences that are similar to the quasi-biennial
variations of solar radiation. The results of calculations of the
radiation fluxes variations periods of stars at the quasi-biennial
scale are given.


\vskip12pt {\it\bf Introduction}
\vskip15pt

The photospheric observations of stars with active atmospheres that
have been conducted regularly  in the optical range since the middle
of the 20th century found low-amplitude variability in some of them,
which is mainly caused by the presence of dark spots on the surfaces
of stars, such as sunspots. Later, at the Mount Wilson Observatory
an extensive program of simultaneous observations of more than a
hundred stars with solar-type chromospheric activity was launched.
These observations (the HK-project) have been performed for over 40
years, from 1965 to the present time [1,2].

In the framework of the HK-project the relationship of the flux in
the centers of the H and K Ca II emission lines (396.8 and 393.4 nm
respectively) to the fluxes in the nearby continuum (400.1 and 390.1
nm respectively) is determined. This relationship  is called the
$S_{Ca II}$ index by authors of the observations. It's known that
the $S_{Ca II}$ index is a good indicator of the chromospheric
activity of the Sun and stars.

Here, we use new data of simultaneous observations of photospheric
and chromospheric fluxes variations of the Sun and 33 solar-type
stars, obtained during the last 20 years at the Lowell (photometric
observations) and Smithsonian observatories (observations of the H
and K Ca II emission lines) of Stanford University during the
HK-project [1]. We use these data for the calculation of the cyclic
activity periods of the  fluxes variations of the Sun and stars.

In this work we confirm the existence of cyclic variations of the
radiation in the quasi-biennial time scale for the majority of the
studied solar-type stars, as well as to determine the periods of
these variations.

\vskip15pt {\it\bf 1. Long-term changes in the activity of
solar-type stars}
\vskip15pt

Solar and stellar activity is a combination of regular manifestation
of characteristic entities in the atmosphere, which are associated
with the release of large amounts of energy, whose frequency and
intensity change cyclically.

In this paper we consider the cyclic variation of Sun's and star's
atmospheres fluxes at 11-year and quasi-biennial time scales. Note
than the duration of the so-called 11-year solar activity varies
from 8 to 15 years according to many years of observations. This
includes regular observations of solar activity during the last one
and half centuries, as well as circumstantial evidence, including
radionuclide and other types of evidence on timescales of 1000
years. The main indices of solar activity that characterize the
study of the full solar disk radiation are the Wolf numbers (the
longest series of observations) and the radio flux at 10,7 cm
wavelength ($F_{10,7}$). Extra-earth atmospheric observations in the
ultraviolet and X-ray ranges of the whole solar disk and of
individual active regions (characterizing the activity at different
altitudes of the atmosphere) were started in the 1960s.
Unfortunately, these observations, which are important for
understanding the nature of the solar activity, are less regular
compared with ground-based observations of Wolf numbers and
$F_{10,7}$. They depend on the operation time of the corresponding
instruments in the circumterrestrial orbit. However, according to
numerous studies of solar activity indices, in particular, flows in
individual lines in short-wave region, all of them correlate well
with Wolf numbers and with a more objective index of activity
$F_{10,7}$.

Observations of stars dive us information only about the magnitude
of radiation fluxes from the full disk. By choosing stars that are
closest in their characteristics to the Sun, we transfer knowledge
about the physical processes in the solar atmosphere to these stars,
which makes it possible to successfully interpret their in radiation
flux variations. Naturally, the Sun and every single solar-like
stars differ in their mass, average density, surface temperature,
the total area of sunspots, their contrast, the dependence of the
temperature of the photosphere and chromosphere on the different
altitudes, sizes and contrast of faculae, and many other parameters.
Analysis of the observations of radiation flux variations of the Sun
and solar-type stars suggests that the structures of the atmospheres
of most of these stars are similar to that of the Sun. The
parameters that differ depend on the mass of a star, its spectral
class, age, chemical composition, and so on. Such an approach makes
it possible to successfully interpret the observations of variations
of radiation fluxes of many solar-type stars, considering the stars
as analogs of the Sun, particulary at different stages of evolution.

For the HK-project stars, periods of 11-year cyclic activity
(according to observations during 40 years [1,2]) change a little
 for one stars and range from 7 to 20 years for various
stars with established activity (about 30\% of the total number of
stars studied in this project).

 By analogy with the detailed study of formations on the Sun,
 the formations on the disk that are responsible for the atmospheric
 activity without flares of solar-type stars include spots and faculae in the photosphere, flocculi
 in the chromosphere, and prominences and coronal mass ejections in the corona.
 Areas where all these phenomena are observed together are known as centers of activity or active regions.

Thus, when analyzing radiation fluxes in the H and K Ca II lines
(which are sensitive to chromospheric activity) and flows from the
entire disc in broadband filters of the photospheric ubvy-system
close to standard UBV-system [1,2] for the Sun and the stars it's
necessary to take into account the contribution from the existing
disk's active regions. Active regions are formed where strong
magnetic fields emerge from underneath the photosphere. In other
words, the different manifestations of the activities of the
atmospheres of the Sun and stars are the result of the magnetic
fields evolution. This involves both global and local magnetic
fields. Their interaction with the magnetized substance in
subphotospheric layers of stars involved in convective motion is
also important. By their size and lifetime active regions vary
greatly. They can be observed from several hours to several mouths.

It was shown that from all solar-type stars that belong to stars of
spectral classes F,G and K of the main sequence on the
Hertzsprung-Russell diagram, a regular cyclicity of chromospheric
activity of the solar type is more common for stars of the late
spectral classes G and K with sufficiently formed subphotospheric
convective zones [3]. These stars rotate relatively slow on their
axes (the rotation period is about 25 - 45 days, in contrast to 3
-10 days for stars with thin subphotospheric convective zones).

The Sun, a star of G2 class, rotates with the period of 25 days. The
rotation period $T_{rot}$ around their axes for a sample of studied
stars, their spectral classes, the quality of their 11-year
cyclicity, and values of respective periods $T^{HK}_{11}$, according
to prior calculations made in [2] are given in table.

We mentioned that all the interpretations of variations in
radiations fluxes of stars are based on the assumption that
solar-type stars (and it is confirmed by all observations) have the
same active regions, which evolve in one or more periods of rotation
around their axes by the same laws that operate on the Sun.

In addition, the time dependence of variations of the radiation flux
from the Sun and stars, which takes into account the rotational
modulation (the time scale is about 1 month), on large scales (of
approximately several years) takes the form of a sine wave with a
period of $T_{11}$ corresponding to the 11-year scale of cyclic
activity.

Based on the example of changes in radiation fluxes in the H and K
Ca II lines for the Sun and stars we see [1] that the amplitude of
the sine wave varies slightly from cycle to cycle. The change in the
amplitude of the light curves of stars for each cycle depends on the
contribution of preexisting and already decayed active regions in
the so-called background radiation in chromospheric lines. An
additional contribution to the amplitude increased in chromospheric
lines is made by higher brightness at the peak of a cycle from the
so-called chromospheric network.

For studies of the Sun, one of the most important problems is to
forecast the cyclical activity, which affects a number of
terrestrial processes.We also consider that the stars of the
HK-project with stable cyclical activity in the 11-year time scale,
established after 40 years of continuous observations, are so
similar in their cyclical activity to the Sun, that for the
prediction of the radiation fluxes from these stars methods that are
used in the practice of solar forecasting are quite applicable
[4,5].

This once again confirms the similarity of the various
manifestations in the evolution of the atmospheric activity of the
Sun and solar-type stars.

\vskip15pt {\it\bf 2. The results of calculations of the variation
periods of the star's fluxes}
\vskip12pt

Recently, in some works the variations of the radiation from the Sun
and stars have been investigated on different time scales with the
help of modern approaches (the application of wavelet analysis to
the observational data). In particular, in [6] the recurrence of
solar activity was analyzed based on observations of sunspots (1750
- 2000) and solar radio-range radiation (1950 - 2000). A distinct
cyclicity of
 the solar activity was revealed with periods of 10 - 11 years and 2
 - 3 years. The authors studied the variations of the radiation of
 solar-type star EI Eri. The period of cyclicity of brightness
 variations of the star has been determined; it is 2.7 years [6], which
 is consistent with our computational results for 33 stars and
 confirms the widespread phenomenon of quasi-biennial variations of
 solar-type stars.

\begin{center}
\centerline{\bf Table}
 \begin{tabular}{|c|c|c|c|c|c|c|c|} \hline
  No & Star   &Sp.class &$T_{rot}$ (day)&Cyclicity,[1] & $T^{HK}_{11}$(yr)&$T_{11}$(yr)&$T_{11}$(yr)\\ \hline
   1 &  Sun   & G2-G4   &    25         &EXCELL        & 10.0             &  10.7      &   2.7     \\ \hline
   2 &HD1835  & G 2.5   &     8         & FAIR         &  9.1             &   9.5      &   3.2     \\ \hline
   3 &HD10476 & K1      &    35         &EXCELL        &  9.6             & 10.0       &    2.8    \\ \hline
   4 &HD13421 & G0      &    17         &UNACT         &   -              & 10.0       &    -      \\ \hline
   5 &HD18256 & F6      &    3          &FAIR          &  6.8             & 6.7        &   3.2     \\ \hline
   6 &HD25998  & F7     &     2         & UNACT        &   -              &   7.1      &   -       \\ \hline
   7 &HD35296 & F8      &    4          &UNACT         &   -              & 10.8        &   -       \\ \hline
   8 &HD39587 & G0      &    5          &UNACT         &   -              & 10.0        &    -      \\ \hline
   9 &HD75332 & F7      &    4          & UNACT        &  -               &   9         &   2.4     \\ \hline
   10 &HD76572 & F6      &   4          &POOR         &  7.1              & 8.5        &    -     \\ \hline
   11 &HD81809  & G2     &   41         &EXCELL       &  8.2              & 8.5        &   2.0      \\ \hline
   12 &HD82885 & G8      &   18         &FAIR         &  7.9              & 8.6        &   -       \\ \hline
   13 &HD103095 & G8     &   31         &EXCELL       &  7.3              & 8.0       &    -      \\ \hline
   14 &HD114710 & F9.5     &    12      &GOOD         &  16.6             & 16         &   2.0     \\ \hline
   15 &HD115383 & G0     &   3          &UNACT        &   -               & 10.3        &    3.5      \\ \hline
   16 &HD115104 & K1     &    18        &GOOD         &  12.4             & 11.8         &   2.7     \\ \hline
   17 &HD120136 & F7    &   4          &POOR          &  11.6              & -        &    3.3      \\ \hline
   18 &HD124570& F6      &    26        &UNACT          &  -              & -         &   2.7     \\ \hline
   19 &HD129333 & G0     &   3          &UNACT        &   -               & 9.0        &    3.2      \\ \hline
   20 &HD131156 & G2     &    6         &UNACT          &  -              & 8.5         &   2.8     \\ \hline
   21 &HD143761 & G0     &   17         &UNACT        &    -              &   -       &    -      \\ \hline
   22 &HD149661 & K2     &    21        &GOOD         &  17.4             & 14.5      &   3.5     \\ \hline
   23 &HD152391 & G7   &     11         &EXCELL        &   10.7           & 10.8        &    2.8      \\ \hline
   24 &HD157856 & F6     &    4        &FAIR          &     -             & 12.9        &   2.6     \\ \hline
   25 &HD158614 & G9     &   34         &UNACT       &     -              & 12.0      &    2.6      \\ \hline
   26 &HD160346 & K3     &    37        &EXCELL         &  7.0             & 8.1         &   2.3     \\ \hline
   27 &HD161239 & G2     &   29         &FAIR        &    5.7              & 6.5        &     -       \\ \hline
   28 &HD182572 & G8     &    41        &UNACT          &  -              & 10.5         &   3.1     \\ \hline
   29 &HD185144 & K0     &   27         &UNACT        &   -               & 6,5         &    2.6     \\ \hline
   30 &HD190007 & K4     &    29        &FAIR         &  10.0             & 11.0         &   2.5     \\ \hline
   31 &HD201091 & K5     &   35         &EXCELL       &   12.0             & 13.7        &    2.8      \\ \hline
   32 &HD201092 & K7     &    38        &GOOD         &  12.4              & 11.7         &   2.5     \\ \hline
   33 &HD203387 & G8     &   -         &UNACT        &   -                &  -          &    2.6      \\ \hline
   34 &HD216385 & F7     &    7        &POOR         &    -               &  7.0         &   2.4     \\ \hline

\end{tabular}
\end{center}

In a recent paper devoted to the quasi-biennial variations (QBV) of
 solar radiation fluxes [7] the importance of the QBV problem studying
was emphasized. It turned out that QBV modulations of total flux of
solar radiation are closely related to various quasi-biennial
variations QBV processes on the Earth, in particular with the QBV of
the Earth's rotation speed, the QBV of the velocity of the
stratospheric wind, etc. The existence of activity of stars on the
quasi-biennial time scale emphasized the fact that solar activity is
a phenomenon that is characteristic of the main sequence stars of
late spectral classes with developed subphotospheric convective
zones. The correct theory of the subphotospheric convection for Sun
and solar-like stars not having been made for the present time [8].

The results of our calculations are shown in the table. We used
observations of radiation fluxes in chromospheric lines (the
$S_{CaII}$ index). We have scanned 33 stars and the Sun's $S_{CaII}$
index light curves using [1] observational data. Then the method of
fast Fourier transform was applied to these scanned data. So we
determined the periods of 11-year cycles (the $T_{11}$) and the
quasi-biennial cyclicity  (the $T_{2}$ period) of chromospheric
activity for the Sun and 33 stars.

A dash in columns with $T_{11}$ and $T_{2}$ means that we have not
identified the periodicity at a given time scale. The table also
presents results of the determination of 11-year periods of stars
$T^{HK}_{11}$, determined for the first time by HK-project
participants and the quality of the cyclicity of the $S_{CaII}$
index variations [1]. One can see a good agreement of these results
with our calculations results of the $T_{11}$ periods.

Note that in some stars with well-determined 11-cyclicity (quality
of the cyclicity is EXCELL or GOOD, as mentioned in table)
variations of fluxes are bimodal near the maximums of the cycle, in
a similar manner to the data for the Sun (according to sun's and
star's observations from [1]).

According to the results of the table, 75\% of the stars of the
sample [1,2] have well-pronounced quasi-biennial fluxes variations,
similar to variation of solar radiation at the same time scale. The
periods of quasi-biennial star's cycles (values $T_{2}$ ) differ in
the range from 2.2 to 3.5.

The preliminary analysis of the QBV in the stars showed that the
duration of this cycle is not constant during one 11-year cycle, as
in the case of QBV of Sun's fluxes. The duration of the QBV of Sun's
fluxes varies on average from 39 months at the beginning of the
11-year cycle to 25 months at the end of it [9].

We compared the data on the photospheric and the chromospheric
emission of HK-project stars according to [1] observations. For
stars with low level of chromospheric activity (Group I, including
the Sun) the correlation is direct. The photospheric emission is
correlated with the $S_{CaII}$ index on time scales of 3 - 20 years.
The other stars (for Group II) are characterized by inversely
directed variations in the photospheric continuum and chromospheric
lines. The correlation coefficients between the photospheric and
chromospheric emission for some stars are high (up to 0,7),
differing by the sign for Groups I and II. One can see that stars
both with the correlation and anticorrelation of activity indexes
corresponding to different atmospheric levels have the QBV of
radiations fluxes.

\vskip15pt {\it\bf Conclusions}
\vskip12pt

A statistical analysis was conducted of observations of the Sun and
33 stars of the HK-project using the fast Fourier transform. From
the spectral analysis of the corresponding changes of radiation
fluxes in Ca II lines we defined the periods of $T_{11}$ 11-year
cyclicity and also revealed a recurrence that similar to the
quasi-biennial solar cyclicity and determined the $T_{2}$ periods.

The table shows that the periods of cycles on the 11-year scale
$T^{HK}_{11}$ identified by observers at Mount Wilson during the
primary treatment of the HK-project observations [2] agree well with
our $T_{11}$ periods.

It can be concluded that the QBV phenomenon of radiation fluxes is
common among solar-type stars. According to our estimates, this QBV
phenomenon is observed in stars about twice as often as
well-determined cyclic activity in the 11-year scale. The periods of
QBV cycles $T_{2}$ for the studied sampling of stars, according to
our calculations, range from 2.2 to 3.5 years.

Acknowledgements. The authors thank the RFBR grant 09-02-01010 for
support of the work.

\vskip12pt {\bf REFERENCES}

1. G.W. Lockwood, B.A. Skif, S.L. Balunas, et al., Astrophys. J.
Suppl., {\bf 171.} P. 260 (2007)

2. S.L. Baliunas , R.A. Donahue R.A.,  W.H. Soon, Astrophys. J.,
{\bf 438.} P. 269 (1995)

3. E.A. Bruevich, M.M. Katsova, and D.D. Socoloff, Astron. Zh., {\bf
78.} P. 48 (2001) [Astron. Rep.,  {\bf 45.} P. 718 (2001)].

4. E.A. Bruevich, Vestn. Mosk. Univ. Fiz. Astron., No. {\bf 6.}, P.
827 (1999)

5. E.A. Bruevich, Astronom. and Astrophys. Transactions, {\bf 23.},
N 2, P.165, (2004), arXiv:1012.5080v1, (2010)

6. Z. Kollath, K. Olah, Astron. and Astrophys., {\bf 501.} P. 695,
(2009)

7. G.S. Ivanov-Kholodnyi and Chertoprud, Sol.-Zemn. Fiz.,  {\bf 2.}
(12), P. 291 (2008)

8.  E.A. Bruevich, I.K.Rozgacheva, arXiv:1012.3693v1 (2010)

9. M.N. Khramova, E.V. Kononovich, and S.A. Krasotkin, Astron.
Vestn., {\bf 36.} P.548 (2002) [Solar Syst. Res., {\bf 36.} P.507
(2002)]

\end{document}